\documentclass[10pt,leqno]{amsart}
\usepackage{graphicx}
\usepackage{indentfirst,csquotes}
\usepackage{multicol, multirow}
\usepackage{makecell}

\usepackage{amssymb,amsthm,amsmath}
\usepackage{xcolor,paralist,hyperref,titlesec,fancyhdr,etoolbox}

\titleformat{\section}[display]{\normalfont\huge\bfseries\centering}{\centering}{10pt}{\Large}
\titlespacing*{\section}{0pt}{0ex}{0ex}

\titleformat{\subsection}{\normalfont\Large\bfseries}{\thesubsection}{1em}{}
\titlespacing*{\subsection}{0pt}{1ex}{1ex}

\hypersetup{ colorlinks=true, linkcolor=black, filecolor=black, urlcolor=black }

\begin{document}
\title{A Defect Classification Framework for AI-Based Software Systems (AI-ODC)} 
\author[Initial Surname]{Mohammed O. Alannsary}
\date{\today}
\email{alannsary@hotmail.com}
\maketitle

\let\thefootnote\relax

\begin{abstract}
Artificial Intelligence has gained a lot of attention recently, it has been utilized in several fields ranging from daily life activities, such as responding to emails and scheduling appointments, to manufacturing and automating work activities. Artificial Intelligence systems are mainly implemented as software solutions, and it is essential to discover and remove software defects to assure its quality using defect analysis which is one of the major activities that contribute to software quality. Despite the proliferation of AI-based systems, current defect analysis models fail to capture their unique attributes. This paper proposes a framework inspired by the Orthogonal Defect Classification (ODC) paradigm and enables defect analysis of Artificial Intelligence systems while recognizing its special attributes and characteristics. This study demonstrated the feasibility of modifying ODC for AI systems to classify its defects. The ODC was adjusted to accommodate the Data, Learning, and Thinking aspects of AI systems which are newly introduced classification dimensions. This adjustment involved the introduction of an additional attribute to the ODC attributes, the incorporation of a new severity level, and the substitution of impact areas with characteristics pertinent to AI systems. The framework was showcased by applying it to a publicly available Machine Learning bug dataset, with results analyzed through one-way and two-way analysis. The case study indicated that defects occurring during the Learning phase were the most prevalent and were significantly linked to high-severity classifications. In contrast, defects identified in the Thinking phase had a disproportionate effect on trustworthiness and accuracy. These findings illustrate AIODC's capability to identify high-risk defect categories and inform focused quality assurance measures. 
\end{abstract} 

\bigskip

\section{Introduction}\label{sec1}

The domain of software quality constitutes a specialized branch within the field of software engineering that rigorously investigates, evaluates, quantifies, and enhances the quality attributes associated with a software product \cite{alannsary2016quality, tian2005software, denning1992software}. In general terms, a software product is regarded as possessing a high level of quality if it is devoid of, or exhibits a minimal number of, issues that originate from the software itself and result in limited adverse effects on its users.

Defect analysis is the branch of software engineering that analyzes and resolves discovered defects while developing the software or using it. When it comes to software quality improvement, defect analysis is  known to be a major contributor. Several defect analysis techniques and methods are available such as defect classification, where found defects are analyzed and grouped based on their characteristics. One of the main and well-known defect classification methods is the Orthogonal Defect Classification (ODC) approach \cite{alannsary2016quality,chillarege1992orthogonal,ma2007web,alannsary2015cloud}.

Implementing defect analysis has been successful in several types of software, such as web applications, Software as a Service (SaaS), and traditional software, yet defect analysis for Artificial Intelligence (AI) systems needs to be addressed due to the additional and different characteristics that AI systems have over traditional systems, web applications, and SaaS.

AI plays a transformative role in improving several fields of life, such as medicine, engineering, economy, social etc. The main advantage of AI is that it helps and aids in data analysis, prediction, and decision making. AI solutions are built mainly using software, therefore it is crucial to evaluate its quality to assure its correctness. One could argue that software written to produce AI does not differ from traditional software, web applications, or SaaS. However, each of the aforementioned software types - including AI systems - has its own attributes and characteristics, and needs to be catered to accordingly \cite{martinez2022software, varma2023bridging, gezici2022systematic, ali2022systematic}.

The quality of the AI system encompasses both the quality of AI as a universal yet specific entity and the quality of AI Platforms (AIP) which are the software and hardware platforms that facilitate its implementation. Therefore, some characteristics were designated for AI quality, some were designated for AIP quality, and some were shared between both AI quality and AI quality \cite{kharchenko2022quality}.

Due to its vast applications, defining attributes and characteristics of an AI system may be troublesome. The main difference between AI systems on one hand and traditional software, web applications, and SaaS on the other is that the former can learn, adapt, and make decisions based on complex unstructured data and provides solutions, which fundamentally emulates human cognitive capabilities \cite{lenarduzzi2021software}, while the latter is based on static rules and predetermined algorithms, works with structured data, and follows established processes. Therefore, there is a need to customize defect analysis techniques to cater for AI systems.

This work proposes AIODC, a defect analysis framework based on the ODC paradigm, and specifically designed for AI systems to aid in classifying its defects. The framework will be predicated on applying the ODC paradigm to AI software while catering for its special characteristics that are different from traditional software, web applications, and SaaS characteristics. On the long-term, AIODC will increase AI system dependability and general quality by assisting in the removal and correction of defects.

Although many studies have examined defect classification in traditional software, cloud computing, and web environments, the distinct defect patterns and quality implications that are specific to AI systems have not been thoroughly investigated. In contrast to conventional software, AI systems present fault vectors associated with their data dependencies, adaptive learning processes, and decision-making logic, which fundamentally change the way defects appear and spread. This highlights the need to not only modify existing classification frameworks such as ODC but also to enhance them with attributes and severity considerations that are specific to AI. By explicitly addressing these gaps, this research establishes AIODC as a customized and context-sensitive instrument for defect analysis in contemporary AI development workflows.

The remainder of this work is organized as follows: the next section presents background and related work. Section 3 discusses ODC, and how it was modified  and/or adapted to different types of software. Section 4 describes the proposed framework. In section 5 a case study is provided to show the applicability of the framework. Section 6 contains the conclusion and prospective.


\section{Background and Related Work}
\label{sec:background}

The procedure for guaranteeing the quality of AI must take into account its unique attributes that diverge from those of conventional software, web-based applications, and SaaS. AI is distinct in that it frequently incorporates novel algorithms and functionalities, undergoes various implementations continuously, and allows developers the flexibility to utilize diverse data sets for the purpose of training the system  \cite{lenarduzzi2021software}.

Consequently, the existing methodologies and frameworks employed to assess and appraise the quality of conventional, web-based software, and SaaS require modification and/or enhancement to suitably recognize the unique attributes associated with AI. Ensuring the reliability of AI is of paramount importance, particularly when decisions are made that may affect human lives \cite{martinez2022software}.

AI is increasingly becoming a crucial asset in the software development life-cycle. This integration signifies a pivotal shift in the approaches utilized for the design, construction, testing, maintenance, and deployment of software. The growing prevalence of AI across diverse sectors is compelling a transition from conventional software development frameworks. Software developers are now able to leverage artificial intelligence to refine processes, minimize errors, and create innovative and intricate software systems utilizing machine learning algorithms, natural language processing, and data analytics. This transition underscores the necessity for organizations to embrace and incorporate AI into various stages of the software development life-cycle (SDLC) \cite{bailey2024impact}. 

Defect analysis for AI software is in its infancy, even though AI has been leveraged to classify and detect defects in several fields of life such as in products, images, and sound. In addition, there has been several attempts to use AI as a tool in the software engineering paradigm, such as code development, testing software, detecting defects, and classifying defects \cite{wang2022software, pandit2022towards, tosun2010ai, pachouly2022systematic, hussain2025leveraging, pandhare2025future, dong2025novel, stradowski2025your, chen2025deep, guzel2025development}. However, there has not been a defect analysis approach developed for AI systems specifically, nor was one introduced in the literature.

Vinayagasundaram and Srivatsa \cite{vinayagasundaram2007software} established metrics to evaluate software quality for AI systems, and stated that AI system components are reusable. In their work, they developed an architecture composed of four layers: task specification, problem solver, domain, and adapter. The main structure is segmented into sub components in a layered manner, allowing for the addition of more layers that can alter the system's behavior. Several metrics were used to measure components quality, such as: depth of inheritance, complexity level, and source code.  

Tao et al. \cite{tao2019testing} elucidated that the unique characteristics of AI software introduce novel challenges and concerns regarding the validation of software or system quality. They discussed the validation of quality pertaining to the functional features of AI software. In addition, they articulated the comprehension of testing AI software in relation to new features and requirements. Moreover, they presented the current classifications of AI software testing and examined different testing strategies. Furthermore, they provided an illustration of test quality evaluation and criteria analysis. Finally, a practical examination of quality validation for an image recognition system was executed through a metamorphic testing technique.

Chillarege et al. \cite{chillarege1992orthogonal} proposed Orthogonal Defect Classification (ODC), a methodology that facilitates in-process feedback to developers by deriving a signature from defects within the development process. Generally, ODC serves as a technique for categorizing defects to identify which phase of the Software Development Life-Cycle requires focus. Defects are categorized based on several dimensions: phase of discovery, defect type, severity, source, trigger, and impact.

Ma and Tian \cite{ma2007web} developed a method for classifying and analyzing web errors through integrating ODC into the web environment. Attributes pertinent to the web context were selected from web access logs to serve as defect types within ODC. This approach facilitated the identification of issues and offered valuable feedback to the web applications development team.

Alannsary and Tian \cite{alannsary2015cloud} proposed a framework for defect analysis tailored for a Cloud-based SaaS. The framework was influenced by the original ODC model and acknowledges the distinctive features of SaaS. It was structured into three phases: data source analysis, classification, and results analysis. The entries from the web server log file are examined to pinpoint errors based on the values of the "Protocol status" field, the detected defects are then classified according to six defect attributes specifically designed for the Cloud context. Successful implementation of one-way and two-way analyses has yielded improved insights into the identified defects and their solutions. A case study was provided to validate the effectiveness of the proposed framework.

Kharchenko et al. \cite{kharchenko2022quality} aimed at establishing and showcasing the implementation of quality models for AI, AIP, and AIS, based on the definition and organization of characteristics. The principles underlying the development of AI quality models and their sequence are thoroughly justified. Approaches for articulating definitions of AIS characteristics, as well as methods for depicting dependencies and hierarchies of characteristics, were discussed. In addition, they proposed definitions and harmonization strategies for the hierarchical relationships among 46 characteristics of AI and AIP. Moreover, the quality models pertaining to AI, AIP, and AIS are detailed, with a focus on the most critical characteristics. Finally, two examples of AIS quality models were elaborated upon. It is worth mentioning that they pointed out that experts have the liberty of including or eliminating characteristics based on the context of the AI system.

Morovati et al. \cite{morovati2023bugs} examined the reproducibility and verifiability of bugs found in Machine Learning (ML) systems, which is part of AI systems, highlighting the key factors associated with each aspect. Subsequently, they delve into the difficulties of creating a benchmark for bugs in ML software systems and introduce a bug dataset to be used as a benchmark called "defect4ML". The dataset meets all the criteria of a standard benchmark, including relevance, reproducibility, fairness, verifiability, and usability. The dataset contained 100 bugs reported by ML developers on platforms such as GitHub and Stack Overflow, utilizing two of the most widely used ML frameworks: TensorFlow and Keras. Cross-platform defect reporting (for instance, GitHub compared to Stack Overflow) poses a risk of duplication, since developers frequently repost issues \cite{zhang2023duplicate}. The process of duplication continues to be difficult in the absence of issue-ID cross-referencing.


\section{Orthogonal Defect Classification (ODC)}
\label{sec:ODC}
Since its first debut, ODC played a noticeable role in defect analysis for traditional software. Several researchers referenced the concept, and some adapted it to different fields. With the introduction of the Web and cloud computing, Software has changed in the way it is delivered and used. Therefore, This section describes how the concept was modified and/or adapted to other types of Software. 

\subsection{Original ODC}
In traditional software and systems, Chillarege et al. \cite{chillarege1992orthogonal} specified a cause effect relationship in ODC. There were two cause attributes: defect type and defect trigger, and several effect attributes such as severity, reliability growth, and impact areas. Table \ref{tab1} shows the cause and effect relationship of the original ODC. 

\begin{table*}
    \centering
\begin{tabular}{m{10em}m{5em}m{12em}}
\cline{1-1}
\cline{3-3}
 CAUSE & \center{\textbf{$\Longleftrightarrow$}} & EFFECT \\
\cline{1-1}
\cline{3-3}
 \multirow{3}{10em}{Defect Type (Development Process)} &  & Severity\\
  & & Impact areas (CUPRIMD)\\
  & & Reliability Growth\\
 \multirow{3}{10em}{Defect Trigger (Verification Process)} & & Defect Density\\
  & & Rework on Fixes\\
  & & Etc.\\
\cline{1-1}
\cline{3-3}
\end{tabular}
\caption{Original ODC cause-effect relationships \cite{chillarege1992orthogonal}, adapted as AIODC’s foundation.}
\label{tab1}
\end{table*}

In addition, in the defect type attribute, defects were classified and associated with a phase in the development processes as shown below.

\begin{description}
    \item [Function] associated with the Design phase.
    \item [Interface] associated with the Low Level Design phase.
    \item [Checking] associated with the Low Level Design or Code phase.
    \item [Assignment] associated with the Code phase.
    \item [Timing/Serialization] associated with the Low Level Design phase.
    \item [Build/Package/Merge] associated with the Library Tools phase.
    \item [Documentation] associated with the Publications phase.
    \item [Algorithm] associated with theLow Level Design phase.
\end{description}

Moreover, defects were classified in regards to the impact areas based on IBM's CUPRIMD \cite{radice1988software}. Table \ref{tab3} shows the attributes and metric areas associated with it.

\begin{table}[h]
\caption{Traditional impact areas (IBM CUPRIMD) \cite{chillarege1992orthogonal} – replaced by AI-specific characteristics in AIODC.}\label{tab3}
\centering

\begin{tabular}{|l|l|}
\hline
 \textbf{Attribute} & \textbf{Metric Areas} \\ \hline
    \multirow{2}{8em}{Capability} & Functionality delivered versus requirements \\
\cline{2-2}
    & Volume of function to deliver \\
\hline
    \multirow{3}{8em}{Usability} & Ease of learning important tasks \\
\cline{2-2}
    & Ease of completing a task \\ 
\cline{2-2}
    & Intuitiveness \\
\hline
    \multirow{3}{8em}{Performance} & Transaction throughput \\ 
\cline{2-2}
    & Response time to enquiry \\
\cline{2-2}
    & Size of machine needed to run the product \\
\hline
    \multirow{4}{8em}{Reliability} & Mean time between failures \\ 
\cline{2-2}
    & Number of defects \\ 
\cline{2-2}
    & Severity of defects \\ 
\cline{2-2}
    & Severity/impact of failures \\
\hline
    \multirow{3}{8em}{Installibility} & Ease of making product available for use \\
\cline{2-2}
    & Time to complete installation \\
\cline{2-2}
    & Skill level needed by installer \\
\hline
    \multirow{2}{8em}{Maintainability} & Ease of problem diagnosis \\ 
\cline{2-2}
    & Ease of fixing problem correctly \\
\hline
    \multirow{3}{8em}{Documentation} & Ease of understanding \\
\cline{2-2}
    & Ease of finding relevant information \\
\cline{2-2}
    & Completeness of information \\
\hline
\end{tabular}
\end{table}

\subsection{Web ODC}
Ma and Tian \cite{ma2007web} adapted the ODC concept to the web environment through identifying web error attributes. The Web ODC attributes are listed below.

\begin{enumerate}
    \item Response code
    \item File type
    \item Referrer type
    \item Time period
\end{enumerate}

\subsection{Cloud ODC}
\begin{figure}
    \centering
	\includegraphics{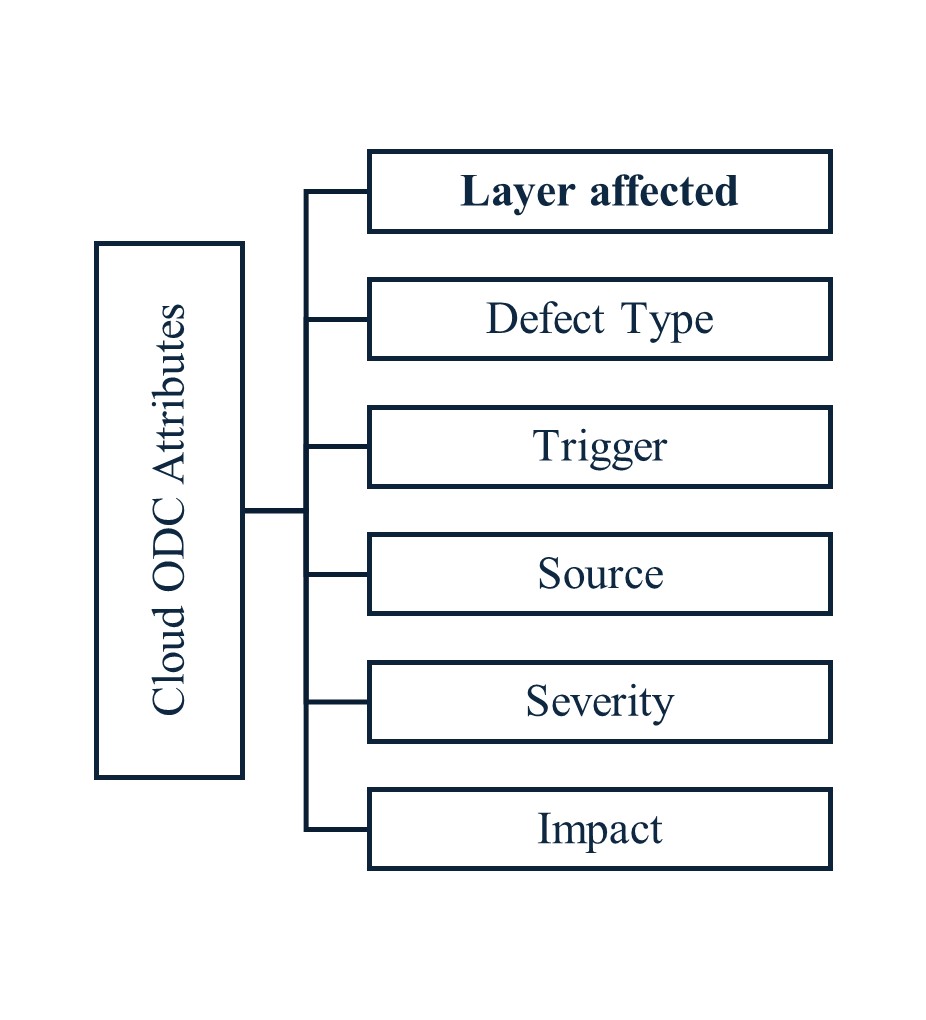}
    \caption{Cloud-ODC attribute schema \cite{alannsary2015cloud} – baseline for AIODC’s extended structure.}
    \label{fig2}
\end{figure}

Due to the special characteristics of SaaS, such as multi-tenancy and isolation, Alannsary and Tian \cite{alannsary2015cloud} adapted the ODC concept to a SaaS running in the Cloud. In the Cloud-ODC framework one attribute (Layer affected) was added to the Original ODC attributes as depicted in Fig. \ref{fig2} including the newly added one which is \textbf{bold}.

In addition, two new defect types (Isolation, IaaS/Paas) were introduced to the original ODC defect types as shown in Table \ref{tab6}, the newly added defect types are \textit{italic}. 

\begin{table} [h]
\caption{Cloud-ODC defect types \cite{alannsary2015cloud} – extended in AIODC with Data/Learning/Thinking dimensions}
\label{tab6}
\centering
\begin{tabular}{|l|l|}
\hline
 \textbf{ODC Version} & \textbf{Defect type} \\ 
\hline
 \multirow{8}{8em}{Original ODC} &  Function.\\
\cline{2-2}
 & Interface \\
\cline{2-2}
 & Checking \\
\cline{2-2}
 & Assignment \\
\cline{2-2}
 & Timing/serialization \\
\cline{2-2}
 & Build/package/merge \\
\cline{2-2}
 & Documentation \\
\cline{2-2}
 & Algorithm \\
    \hline
 \multirow{2}{8em}{Cloud-ODC} & \textit{Isolation}\\
\cline{2-2}
 & \textit{IaaS/PaaS}\\
\hline
\end{tabular}

\end{table}


\section{Proposed Framework}
\label{sec:framework}

Defining attributes and characteristics of an AI systems is essential to allow classification of its defects. However, known software/system characteristics are not sufficient due to the special nature of AI systems. Several published literature discussed AI quality and AIP quality models such as \cite{ali2022systematic, gezici2022systematic}. 

An AI system is different from other systems in the sense that it emulates human cognitive capabilities, it can learn, adapt, make decisions, and provide solutions based on complex unstructured data. To utilize the classification paradigm of the ODC model in AI system development projects, it is essential to undertake particular actions that cater to the distinctive features of AI. Therefore, the original ODC needs to be modified to be adapted to the defect analysis process of AI systems.

The AIODC classification process employs a systematic and repeatable methodology to guarantee consistency. Initially, each reported defect is scrutinized to ascertain whether it relates to AI-specific behavior by categorizing it into one of four AI attribute categories: Data (issues within training/testing datasets), Learning (errors during model training), Thinking (mistakes in inference or decision-making), or Not Related (general software defects). Each category is accompanied by explicit examples, such as mislabeled training data for Data defects, unstable convergence in Learning defects, or incorrect decision thresholds in Thinking defects. Severity ratings are allocated based on established decision rules that take into account the application domain, potential downstream effects, and the reversibility of outcomes, ensuring that “Catastrophic” severity is designated for defects with a significant risk of irreversible damage in critical domains. Ultimately, the impact mapping process utilizes Kharchenko’s AI/AIP quality model, with annotators directed by a rubric that connects defect types to quality characteristics grounded in functional dependencies and observed system behavior.

\begin{figure}
    \centering
	\includegraphics{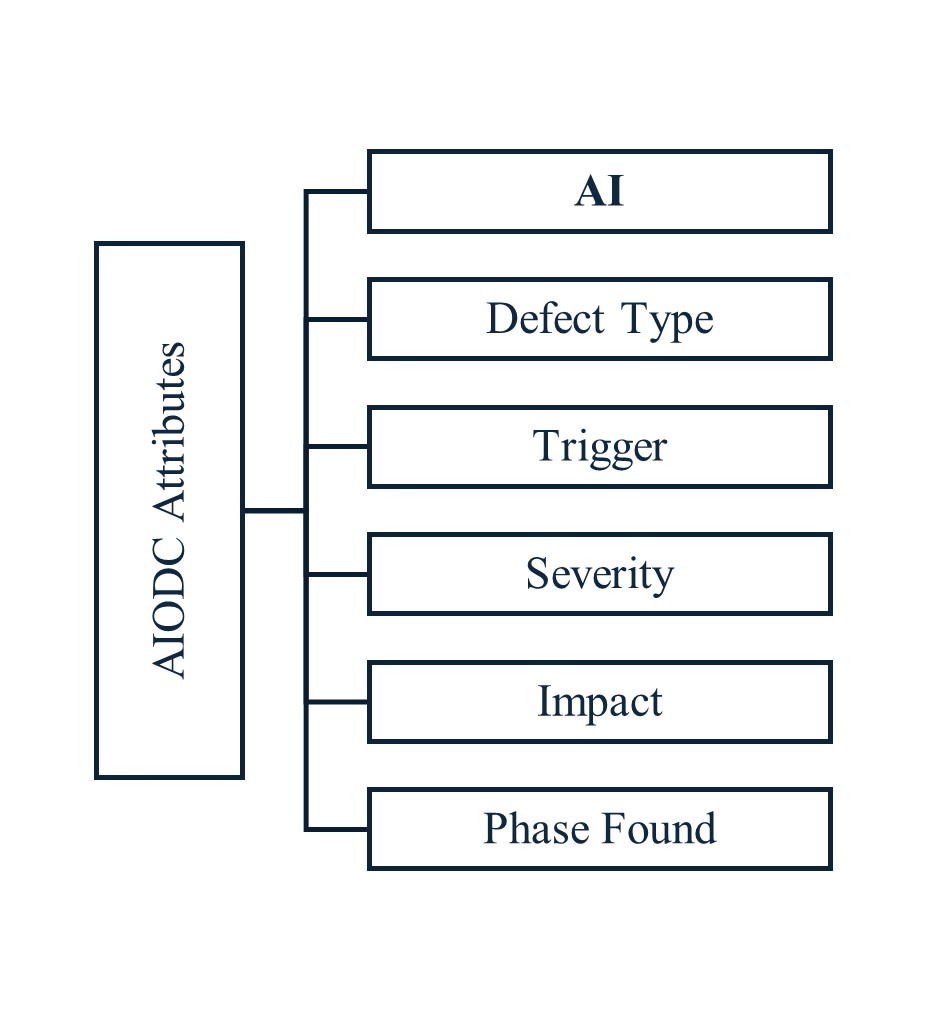}
    \caption{AIODC’s attribute framework – integrating novel AI dimension with ODC core.}
    \label{fig3}
\end{figure}

First, the list of attributes is modified by introducing a new attribute called AI. The new attribute allows classifying defects based on Data, Learning, Thinking, and Not Related as explained in Table \ref{tab11}. The attribute allows classifying defects based on the Data used to train the AI, the AI's learning process, and the AI's training process, or not related to AI respectively. Fig. \ref{fig3} depicts the suggested AIODC attributes including the newly added one which is \textbf{bold}.

\begin{table} [h]
\caption{AIODC’s novel AI attribute – classifying defects by Data, Learning, Thinking, or Not Related phases}
\label{tab11}
    \centering
\begin{tabular}{|l|l|}
\hline
 \textbf{Classification} & \textbf{Description}\\ 
\hline
 Data & Issues with training / testing data.\\
    \hline
 Learning & Faults in the AI model training process.\\
    \hline
 Thinking & Faults in inference, logic, or decision making.\\
    \hline
 Not Related & Defects unrelated to AI logic or behavior.\\
\hline
\end{tabular}
\end{table}

\begin{figure}
    \centering
	\includegraphics{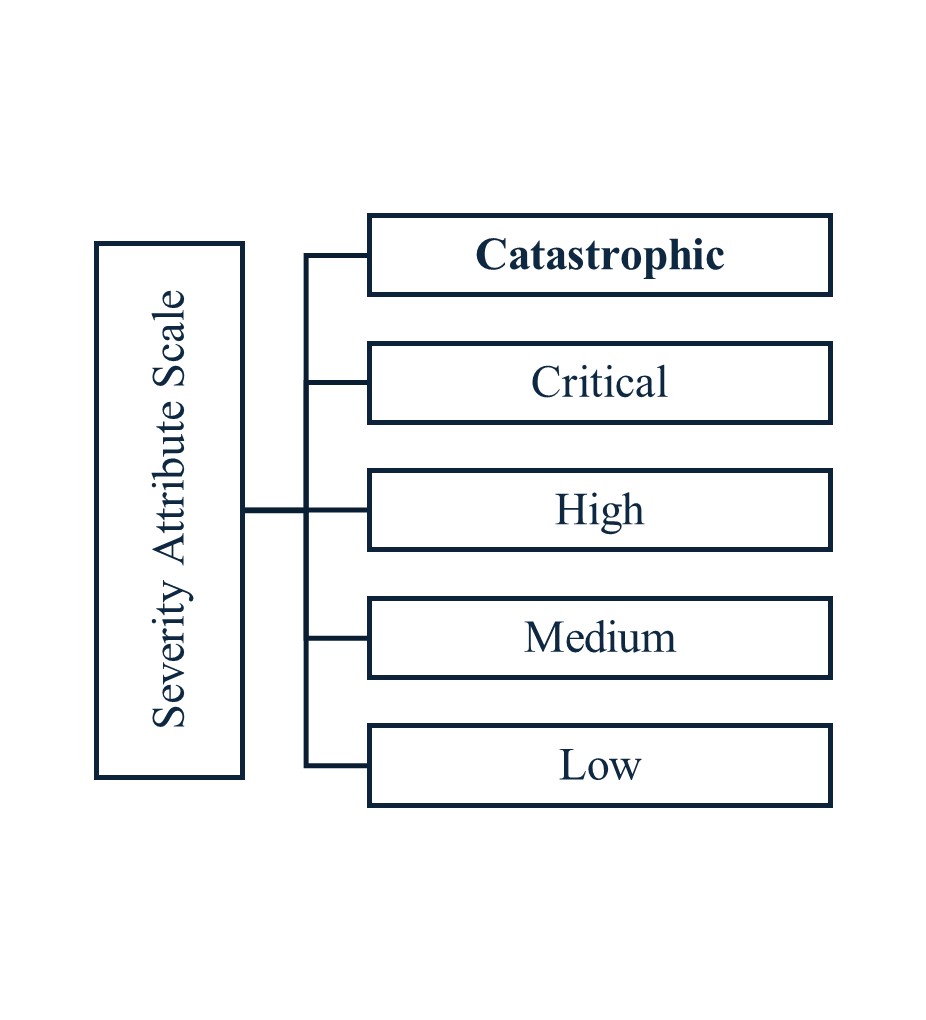}
    \caption{AIODC’s 5-tier severity scale – with domain-adaptive "Catastrophic" level for AI risks.}
    \label{fig4}
\end{figure}

Second, the severity attribute is changed from a four scale measure to a five scale measure by adding a new severity level named "Catastrophic". The severity attribute in the original ODC is measured on a scale from one to four (Critical, High, Medium, and Low), since defects in an AI systems may be catastrophic due to the fact that they may be used in numerous significant applications including ones related to human health monitoring, and illness diagnosing \cite{Lenarduzzi2023health}. Therefore, the severity attribute is modified to be measured on a one to five scale (Catastrophic, Critical, High, Medium, and Low). Fig. \ref{fig4} depicts the suggested Severity attribute scale including the newly added level which is \textbf{bold}. To address context-sensitive risk profiles in AI systems, severity classifications encompass three key factors:

\begin{enumerate}
    \item Application Criticality:
    \begin{itemize}
        \item Catastrophic/Critical: Systems that affect human safety (healthcare, autonomous vehicles).
        \item High/Medium: Enterprise systems (finance, security)
        \item Low: Non-critical applications (entertainment, productivity)
    \end{itemize}
    \item Harm Reversibility:
    \begin{itemize}
        \item Irreversible harm, such as: fatal misdiagnosis (Catastrophic)
        \item Reversible errors, such as:transient data corruption (High/Medium)
    \end{itemize}
    \item Failure Scope:
    \begin{itemize}
        \item Systemic failures, such as: poisoned training data (increase severity)
        \item Localized errors, such as: single inference fault (allow for lower severity)
    \end{itemize}
\end{enumerate}

This methodology is consistent with recognized risk management standards, including ISO 14971:2019 for medical devices \cite{iso14971_2019}, which highlights severity levels that depend on context. Similarly, harm taxonomies specific to AI (for instance, irreversible systemic failures) are informed by fundamental safety literature \cite{amodei2016concrete}.

\begin{figure*}
    \centering
	\includegraphics{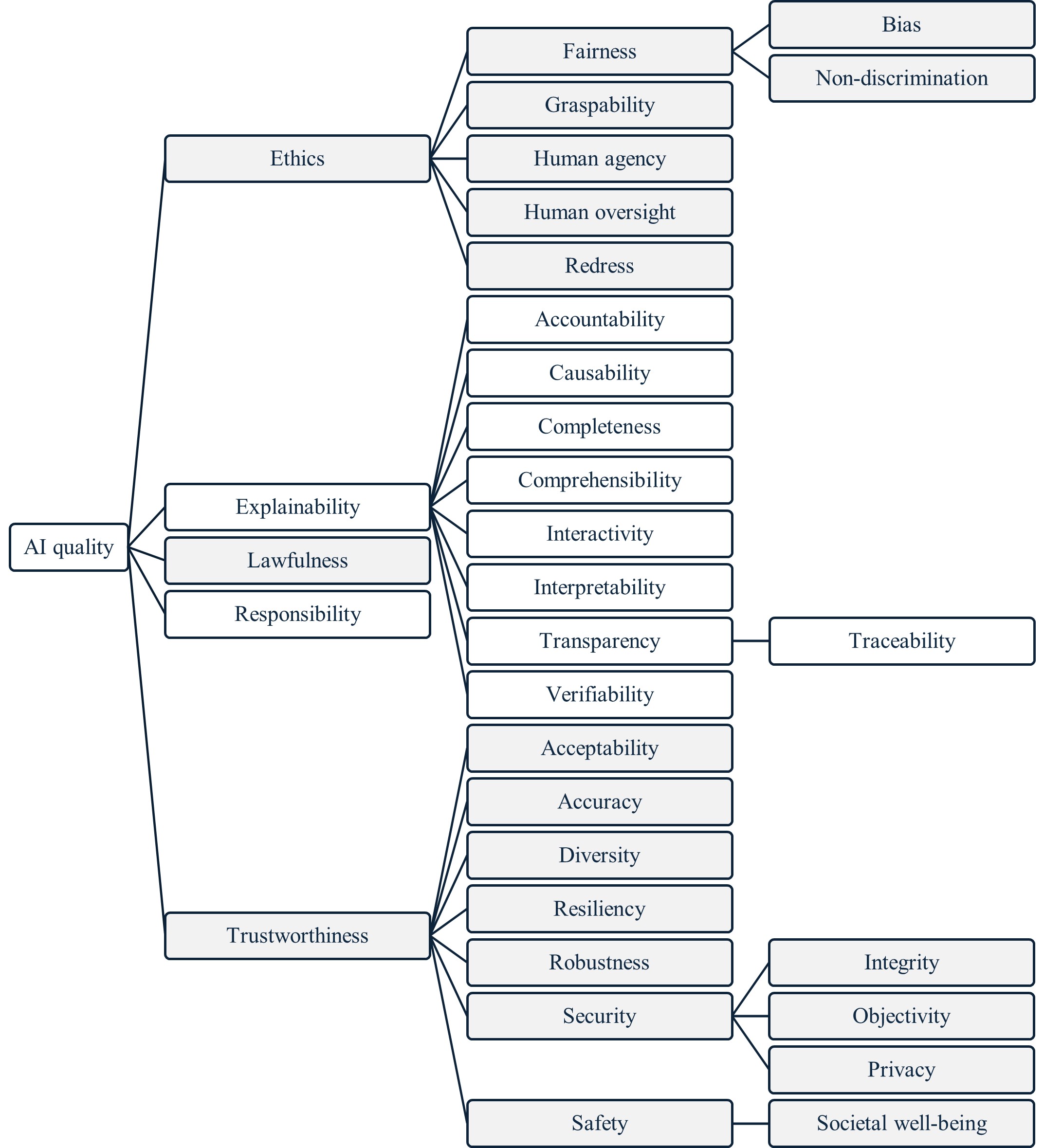}
    \caption{AI quality characteristics (Kharchenko et al. \cite{kharchenko2022quality}) – foundation for AIODC impact mapping.}
    \label{fig5}
\end{figure*}

\begin{figure*}
    \centering
	\includegraphics{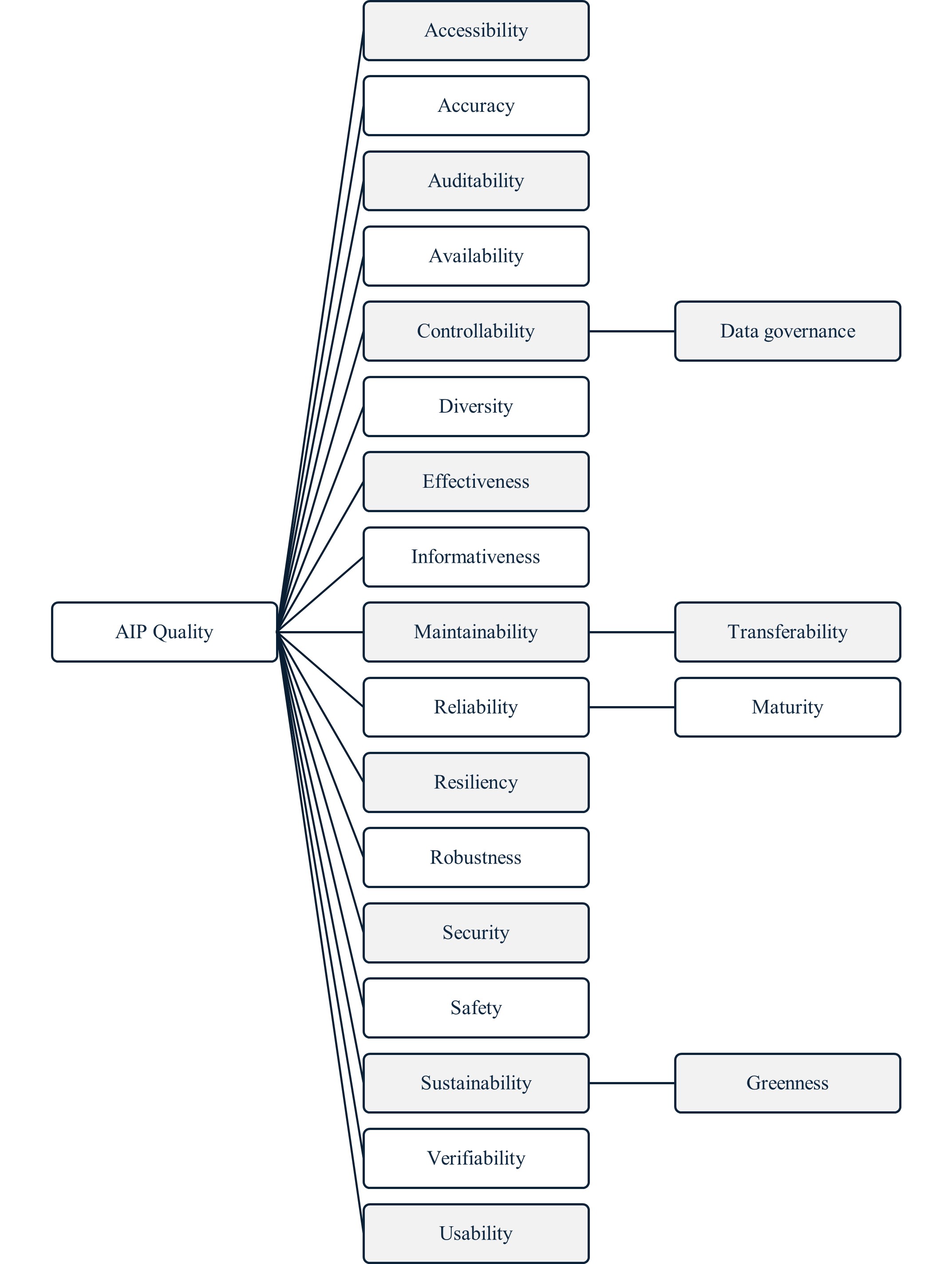}
    \caption{AIP quality characteristics (Kharchenko et al. \cite{kharchenko2022quality}) – foundation for AIODC impact mapping.}
    \label{fig6}
\end{figure*}

Finally, the impact areas attribute is modified to accommodate the new characteristics of AI system quality. As explained previously, Kharchenko et al. \cite{kharchenko2022quality} developed a quality model and suggested a list of characteristics that are appropriate for AI systems, where the quality of an AI system is composed of the quality of AI and the quality of AIP. The characteristics were distributed among three layers, so if characteristic C1 depends on characteristic C2, then C2 will be in the next lower layer that C1 is in. In addition, some characteristics were assigned to AI quality, some assigned to AIP quality, and some were shared between both. Fig. \ref{fig5} depicts AI quality characteristic, while Fig. \ref{fig6} depicts AIP quality characteristic. 


\section{Case Study}

To test the applicability and suitability of the AIODC framework for AI systems, the benchmark dataset created by Morovati et al. \cite{morovati2023bugs} is utilized, comprising 100 ML defects sourced from TensorFlow/Keras on GitHub and Stack Overflow. The selection of Keras defects from GitHub (n=42) was based on three primary criteria:
\begin{enumerate}
    \item Highest defect count: 42 compared to TensorFlow’s 20 on GitHub;
    \item Uniformity of platform: to prevent duplication in cross-platform reporting;
    \item Consistency within the framework: to minimize confounding variables arising from differences between frameworks.
\end{enumerate}

Therefore, defects reported for TensorFlow or on Stack Overflow were excluded to prevent duplication bais.

As mentioned above, the dataset contains 100 bugs reported on two platforms: GitHub and Stack Overflow, for two widely used ML frameworks: TensorFlow and Keras. In this study, defects reported on GitHub for the Keras framework will classified based on the proposed AIODC framework. Such decision was made based on the number of defects that each framework has on each platform. Table \ref{tab10} shows the number of defects for each framework on both platforms.

\begin{table} [h]
\caption{Defect distribution in benchmark dataset \cite{morovati2023bugs} –Keras/GitHub selected (n=42) for case study due to maximal unique defects}
\label{tab10}
\centering
\begin{tabular}{|l|c|c|}
\hline
 \textbf{Platform} & \textbf{TensorFlow} & \textbf{Keras}\\ 
\hline
 GitHub & 20 & 42\\ 
\hline
Stack Overflow & 3 & 35\\ 
\hline
\end{tabular}
\end{table}

\subsection{Classification Methodology}
To classify defects to implement AIODC, Two separate annotators examined defects according to predefined guidelines:
\begin{enumerate}
     \item Training: A two-hour session featuring sample defect mappings.
    \item Procedure:
        \begin{itemize}
            \item Annotators evaluated descriptions of GitHub issues
            \item Designated AI attributes (Data/Learning/Thinking/Not Applicable)
            \item Assigned severity ratings (Catastrophic-Low)
        \end{itemize} 
    \item Dispute Resolution: A third expert evaluated the conflicting labels
    \item Reliability Assessment: Cohen's Kappa = 0.82 (indicating substantial agreement)
\end{enumerate}

\subsection{Bug Classification based on the AI attribute}
The number of defects reported for the Keras framework on GitHub is 42 defects. Classifying defects based on the AI attribute resulted in having 2 defects related to Data, 18 related to the Learning process of the AI framework, 14 related to the Thinking process of the AI framework, and 8 defects not related to AI. Classification results are shown in Table \ref{tab14}.

\begin{table*}
\caption{AIODC classification of unique Keras defects (GitHub subset, n=42)}
\label{tab14}
    \centering
\begin{tabular}{|l|c|c|c|c|}
\hline
    \textbf{Defect Description} & \textbf{Count} & \textbf{AI Attribute} & \textbf{Total} & \textbf{\%} \\ 
\hline
    Deprecated API & 3 & Not Related & \multirow{4}{*}{8} & \multirow{4}{*}{19.05\%}\\
    \cline{1-3}
    Missing API call & 2 & Not Related &  &  \\
    \cline{1-3}
    Missing argument scoping & 1 & Not Related &  &  \\
    \cline{1-3}
    Wrong API usage & 2 & Not Related  &  & \\
    \hline
    Missing dense layer & 1 & Think & \multirow{6}{*}{14} & \multirow{6}{*}{33.33\%}\\
    \cline{1-3}
    Suboptimal network structure & 4 & Think & & \\
    \cline{1-3}
    Wrong size for convolutional layer & 1 & Think &  & \\
    \cline{1-3}
    Wrong layer type & 2 & Think &  & \\
    \cline{1-3}
    Wrong network architecture & 3 & Think &  & \\
    \cline{1-3}
    Wrong type of activation function & 3 & Think &  & \\
    \hline
    Wrong tensor shape & 1 & Data & \multirow{2}{*}{2} & \multirow{2}{*}{4.76\%}\\
    \cline{1-3}
    Missing pre processing step & 1 & Data &  & \\
    \hline
    Suboptimal batch size & 4 & Learn & \multirow{5}{*}{18} & \multirow{5}{*}{42.86\%}\\
    \cline{1-3}
    Suboptimal number of epochs & 4 & Learn &  & \\
    \cline{1-3}
    Wrong loss function calculation & 1 & Learn &  & \\
    \cline{1-3}
    Wrong optimization function & 4 & Learn & & \\
    \cline{1-3}
    Wrong selection of loss function & 5 & Learn &  & \\
    \hline
\end{tabular}
\end{table*}

\subsection{Bug Classification based on the Severity attribute}

The severity of defects was categorized based on domain-adaptive thresholds outlined in Section 4, which evaluate:

\begin{enumerate}
    \item Application Criticality: Keras (general ML development) was classified as medium-criticality (non-safety-critical).
    \item Harm Reversibility: Irreversible errors (such as model collapse) heightened the severity level.
    \item Failure Scope: Systemic defects (for instance, hyperparameter corruption) were given precedence over localized problems.
\end{enumerate}

Therefore, the revised Severity Assignment Logic is:
\begin{enumerate}
    \item Catastrophic (5): Defects that result in irreversible systemic damage (e.g., "incorrect loss function" causes permanent divergence of the model).
    \item Critical (4): Defects that lead to recoverable yet significant outcomes (e.g., "incorrect tensor shape" causes failure in training; can be rectified through data correction).
    \item High (3) to Low (1): Localized or transient defects (e.g., "suboptimal epochs" causes decrease in accuracy; can be resolved through retuning).
\end{enumerate}

Severity classifications were aligned with the domain-adaptive standards outlined in Section 4, which encompass criticality levels inspired by ISO 14971 \cite{iso14971_2019} and the AI failure modes detailed in \cite{amodei2016concrete}. In the case of Keras, categorized as medium-criticality, defects were classified as 'Catastrophic' solely under the following conditions:

\begin{itemize}
    \item Irreversible systemic damage,
    \item Equivalence of life/financial risk (according to \cite{iso14971_2019} Annex C),
    \item Uncontainable propagation of errors (as specified in \cite{amodei2016concrete} Sec 3.1).
\end{itemize}

The application of these criteria to defects in Keras produced the classification shown in Table \ref{tab13}.

\begin{table} [h]
\caption{Severity classification – validating domain-adaptive thresholds (n=42)}
\label{tab13}
\centering
\begin{tabular}{|l|c|c|}
\hline
    \textbf{Severity level} & \textbf{Defects (n=42)} & \textbf{\%}  \\ 
\hline
    Catastrophic & 9 & 21.43\%\\
\hline
    Critical & 10 & 23.81\%\\
\hline
    High & 12 & 28.57\%\\
\hline
    Medium & 11 & 26.19\%\\
\hline
    Low & 0 & 0.00\%\\
    \hline
\end{tabular}

\end{table}

\subsection{Bug Classification based on the Impact attribute}

Expanding upon the classification of defects, we subsequently examine the manner in which AI defects adversely affect the fundamental quality characteristics of AI systems and their associated platforms. Unlike traditional defect analysis, this approach isolates AI-specific failure vectors through quality-attribute-centric mapping. Although categorization serves to pinpoint the origins of failures (Data/Learning/Thinking/Not Related), effective remediation necessitates a comprehension of the specific quality characteristic that these defects undermine. This section correlates defects identified in the case study with empirically observed degradations in the quality characteristics of AI/AIP (following Kharchenko's quality model \cite{kharchenko2022quality} for characteristic definitions), thereby uncovering deterministic patterns between defect types and impacted attributes. These consistent failure signatures facilitate accurate localization of quality risks — thereby affirming AIODC's core proposition that the sources of defects dictate impact profiles across functionally orthogonal quality dimensions. Table \ref{tab12} shows the classification of found defects based on their impact on the Kharchenko AI/AIP Characteristics.

\begin{table*} 
\caption{Impact mapping of Keras defects to AI/AIP quality characteristics \cite{kharchenko2022quality}}
\label{tab12}
    \centering
\begin{tabular}{|l|c|l|l|c|}
\hline
    \multirow{2}{*}{\textbf{AI Defect}} & \multirow{2}{*}{\textbf{\makecell{Quality\\Model}}} & \multicolumn{3}{c|}{\textbf{Impacted characteristic}}  \\
    \cline{3-5}
    & & \textbf{Layer1} & {Layer2} & {Layer3} \\ 
\hline
    Deprecated API & AIP & Maintainability & = & =\\
\hline
    Missing API call & AIP & Reliability & = & = \\
\hline
    Missing argument scoping & AI & Security & Integrity & = \\
\hline
    Wrong API usage & AIP & Accuracy  & = & = \\
\hline
    \multirow{2}{*}{Missing dense layer} & AI & Trustworthiness & Accuracy & = \\
    \cline{2-5}
    & AIP & Accuracy & = & = \\
\hline
    Suboptimal network structure & AI & Effectiveness & = & = \\
\hline
    \multirow{2}{*}{Wrong size for convolutional layer} & AI & Trustworthiness & Robustness & = \\
    \cline{2-5}
    & AIP & Robustness & = & = \\
\hline
    \multirow{2}{*}{Wrong layer type} & AI & Trustworthiness & Accuracy & = \\
    \cline{2-5}
    & AIP & Accuracy & = & = \\
\hline
    \multirow{3}{*}{Wrong network architecture} & AI & Trustworthiness & Accuracy & = \\
    \cline{2-5}
    & AI & Explainability & Completeness & = \\
    \cline{2-5}
    & AIP & Accuracy & = & = \\
\hline
    \multirow{2}{*}{Wrong type of activation function} & AI & Trustworthiness & Accuracy & = \\
    \cline{2-5}
    & AIP & Accuracy & = & = \\
\hline
    Wrong tensor shape & AIP & Reliability & = & = \\
\hline
    \multirow{2}{*}{Missing pre processing step} & AI & Trustworthiness & Robustness & = \\
    \cline{2-5}
    & AIP & Robustness & = & = \\
\hline
    Suboptimal batch size & AI & Effectiveness & = & = \\
\hline
    \multirow{3}{*}{Suboptimal number of epochs} & AI & Trustworthiness & Accuracy & = \\
    \cline{2-5}
    & AIP & Accuracy & = & = \\
    \cline{2-5}
    & AIP & Effectiveness & = & = \\
\hline
    \multirow{2}{*}{Wrong loss function calculation} & AI & Trustworthiness & Accuracy & = \\
    \cline{2-5}
    & AIP & Accuracy & & \\
\hline
    Wrong optimization function & AI & Effectiveness & = & = \\
\hline
    \multirow{4}{*}{Wrong selection of loss function} & AI & Trustworthiness & Accuracy & = \\
    \cline{2-5}
    & AI & Trustworthiness & Robustness & = \\
    \cline{2-5}
    & AIP & Accuracy & = & = \\
    \cline{2-5}
    & AIP & Robustness & = & = \\
    \hline
\end{tabular}
\end{table*}

The classification of defects in relation to AI/AIP quality characteristics shown in Table \ref{tab12} was performed through a two-phase expert evaluation procedure. Annotators employed a specified rubric that correlated defect types with Kharchenko’s hierarchical quality characteristics, taking into account both the functional role of the defect's location (such as model architecture and training loop) and the type of failure (for instance, robustness degradation and accuracy loss). In instances of uncertainty, annotators referred to documentation and discussions from issue threads to gain a clearer understanding of the defect's context prior to finalizing the labels. This systematic mapping process minimizes subjectivity and improves reproducibility, thereby ensuring that subsequent researchers can utilize AIODC on other datasets with similar consistency.

Analysis indicates clear failure patterns across AI defect categories: Defects originating in the Learning phase predominantly degrade operational efficiency metrics (such as computational resource usage and training duration), while those emerging from the Thinking phase primarily compromise system integrity outcomes (including decision reliability and output consistency). This illustrates how defect genesis in AI systems dictates failure modes across functionally orthogonal impact domains.

This division illustrates the architectural distinction between the training and inference subsystems. Learning defects present themselves as resource-related issues, whereas Thinking defects arise as behavioral irregularities.

\subsection{Two-Way Classification}

\begin{figure*}
    \centering
	\includegraphics{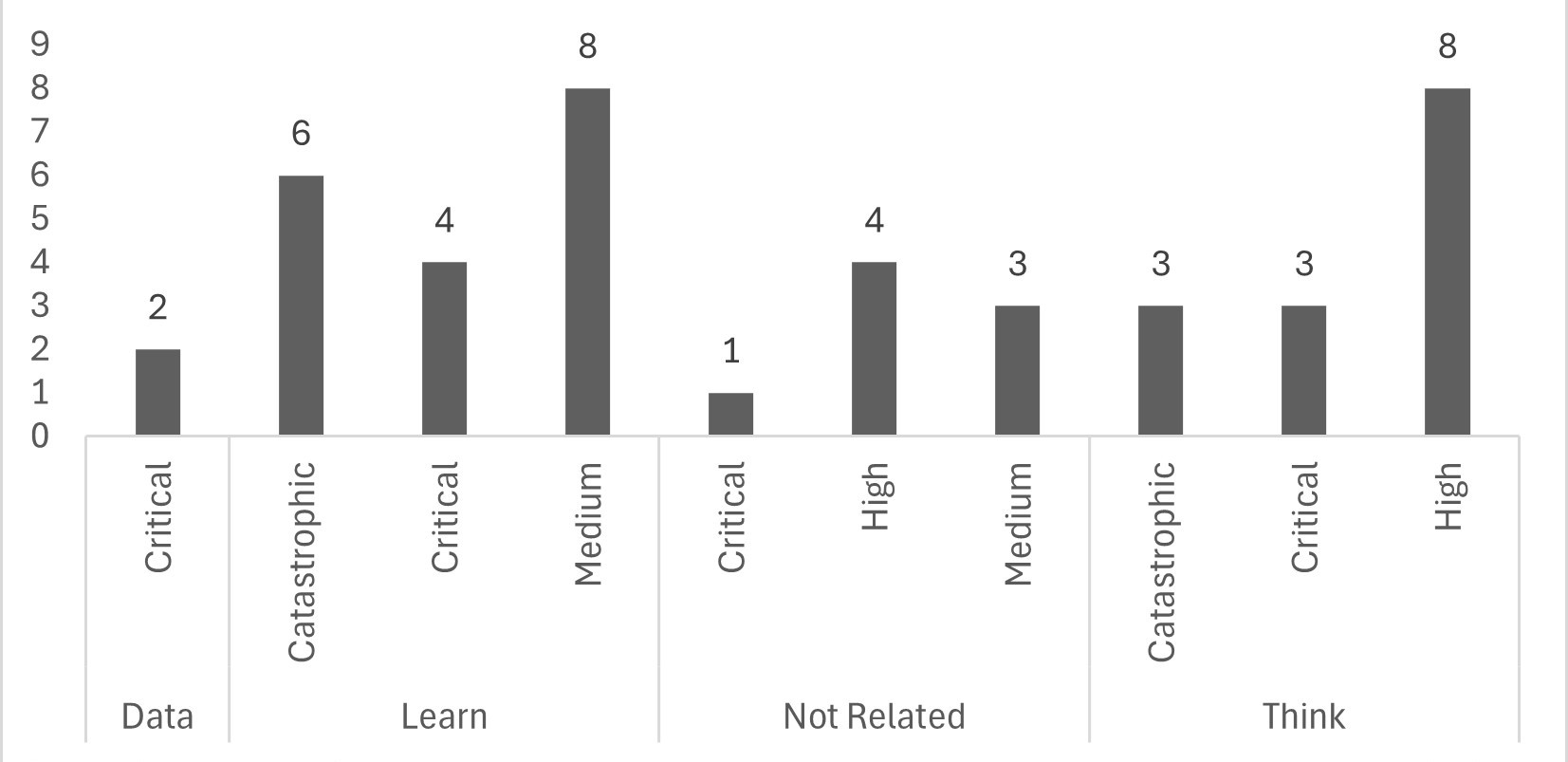}
    \caption{AI/Severity correlation – revealing high-risk defect clusters (e.g., Learning + Catastrophic).}
    \label{fig9}
\end{figure*}

While analyzing defects classified using one attribute (one-way classification) can be insightful. However, it may not adequately inform the development team about defects and the associated high-risk areas. Implementing two-way classification will enhance the understanding of the observed defects, which may lead to more informed decisions. Therefore, by utilizing two-way classification, it becomes feasible to concentrate on rectifying high-priority defects, ultimately improving the reliability and quality of the AI system. Two way classification was implemented on the AI and Severity attributes, results are depicted in Fig. \ref{fig9}.

\section{Discussion}

The findings of this research have significant implications for the practices of quality assurance in AI software. By establishing a clear connection between the origins of defects and their severity outcomes, as well as the quality characteristics affected, AIODC empowers development teams to focus on defects that present the greatest operational and safety hazards. For instance, defects categorized under Thinking that compromise trustworthiness or accuracy may require urgent intervention in healthcare AI, whereas defects arising during the Learning phase that reduce efficiency might be prioritized in large-scale industrial AI applications where computational expenses are paramount. Additionally, the organized AIODC framework could be incorporated into continuous integration workflows, facilitating real-time assessments of defect risks and proactive quality management in AI development.

From a methodological standpoint, the introduction of a new AI attribute (Data, Learning, Thinking, Not Related) alongside a five-tier severity scale offers a more comprehensive and nuanced defect profile compared to current ODC adaptations. The capacity to distinguish AI-specific challenges from general software defects enables more focused debugging approaches and supports the development of specialized test cases for data management, model training, and inference processes. The identified correlation between AI attribute categories and severity levels strengthens the hypothesis that the type of defect serves as an indicator of potential harm, which is essential in risk-based testing and quality strategy formulation.

The classification of impact utilizing Kharchenko’s AI/AIP quality model has also uncovered persistent failure patterns: defects related to Learning were found to diminish both efficiency and robustness, whereas defects associated with Thinking had a disproportionate effect on trustworthiness and accuracy. This observation reinforces the architectural separation between the training and inference subsystems of AI and indicates that measures for defect prevention ought to be tailored to each subsystem. For example, tools for automated training monitoring may prove to be more effective in identifying anomalies during the Learning phase, while audits focused on explainability could more effectively address irregularities occurring in the Thinking phase.

From a theoretical perspective, this research adds to the developing domain of Software Engineering for AI by transforming quality models into a practical framework for defect analysis. Although previous studies have suggested quality metrics or general defect taxonomies, AIODC serves to connect abstract quality attributes with tangible workflows for defect classification. This linkage is crucial for enhancing quality assurance processes in AI initiatives, where the detection of defects is complicated by non-deterministic behaviors and evolving models.

Nevertheless, the case study’s concentration on a singular dataset and framework (specifically, Keras issues on GitHub) defects limits generalizability. This intentional design mitigated cross-platform redundancy risks, as mentioned above, yet it constrains insights into cross-framework defect patterns. Future work will:
\begin{enumerate}
    \item Implement deduplication techniques, such as issue-ID matching \cite{zhang2023duplicate} to facilitate analysis across multiple platforms.
    \item Validate AIODC on TensorFlow defects to assess framework-agnostic performance.
\end{enumerate}


\section{Conclusion and Prospective}
\label{sec:conclusion}

Artificial Intelligence systems are predominantly deployed as critical software solutions across a wide range of domains, making the identification and resolution of software defects essential to ensure system quality. Considering the distinct attributes of AI systems – including their ability to learn, dependency on data, and decision-making logic – conventional defect analysis methods are inadequate in addressing the practical challenges they present.

This study presented AIODC, a tailored defect classification framework inspired by the Orthogonal Defect Classification (ODC) model, specifically tailored to address the complexity of AI systems. The framework enhances ODC by integrating a novel attribute focusing on AI-related dimensions (Data, Learning, and Thinking), broadening the severity scale to introduce “Catastrophic” level, and adapting impact areas to reflect the quality characteristics of AI systems. These improvements offer a more refined and context-aware approach for classifying defects in AI environments.

The applicability of the framework was illustrated through a case study that utilized a publicly accessible dataset of defects from a machine learning platform. Results from one-way and two-way classification illustrated the feasibility of employing AIODC to reveal insights regarding the origins of defects and patterns of severity, thereby supporting more effective debugging and quality assurance in AI development workflows.

A limitation of this study is its present validation that is centered on machine learning systems (Keras). Subsequent research will evaluate natural language processing systems (NLP's), real-time artificial intelligence (e.g. autonomous vehicle frameworks), and Non-neural network methodologies (e.g. Bayesian artificial intelligence).

Future work will focus on four key directions:
\begin{enumerate}
    \item Framework validation and extension:
    \begin{itemize}
        \item Apply AIODC to larger and more diverse datasets to validate its scalability and generalizability.
        \item Validate AIODC performance in industrial scale systems.
    \end{itemize}
    \item Automation and scalability
    \begin{itemize}
        \item Incorporate AIODC with Continues Integration/Continues Delivery and Deployment (CI/CD) pipelines for real-time defect classification.
        \item Leverage Large Language Models (LLMs) to automate defect classification and reduce manual overhead.
    \end{itemize}
    \item Operational enhancements
    \begin{itemize}
        \item Integrate AIODC into automated issue-tracking systems for longitudinal defect trend analysis.
        \item Establish standardized inter-rater reliability protocols.
    \end{itemize}
    \item Empirical enrichment
    \begin{itemize}
        \item  Utilize larger datasets to facilitate more robust statistical inferences of defect-impact relationships.
        \item Quantify risk reduction through AIODC-guided quality interventions.
    \end{itemize}
\end{enumerate}


\bibliographystyle{acm}
\bibliography{bibliography}

\end{document}